\documentclass{elsart}

\begin{document}

\begin{frontmatter}

\title{Multi-band Wigner Function Formulation of Quantum Transport}

\author{Mehmet Burcin Unlu, Bernard Rosen, and Hong-Liang Cui}

\address{Department of Physics and Engineering Physics, Stevens Institute
of Technology, Hoboken, New Jersey 07030}

\author{Peiji Zhao}

\address{Department of Electrical and Computer Engineering, North Carolina
State University, Raleigh, North Carolina 27695}

\maketitle

\begin{abstract}
A Wigner function representation of multi-band quantum transport
theory is developed in this paper. The equations are derived using
non-equilibrium Green's function formulation with the generalized
Kadanoff-Baym ansatz  and the multi-band $\bf{k.p}$ Hamiltonian
including spin. The results are applied to a two-band resonant
inter-band tunneling structure.
\end{abstract}

\end{frontmatter}

\section{Introduction}

The single band approximation \cite{key-25} is the most often used
approach in quantum device models. In this approximation, inter-band
processes in the structure are ignored and the boundary conditions
for the model are oversimplified. The single band electron transport
models have been applied to large band-gap semiconductor heterostructures
(e.g. AlAs/GaAs/AlAs).

Multi-band quantum transport has attracted attention due to the
existence of various inter-band tunneling structures and
especially resonant inter-band tunneling structures (RITS). It is
possible to achieve multiple negative differential regions (NDR)
in these structures and they are shown to exhibit high
peak-to-valley current voltage characteristics. RITS's are based
upon the type-I, the type-II staggered and the type-II broken-gap
band alignments. Recently \cite{key-5}, it was theoretically shown
that a type-II staggered band-gap resonant tunneling diode can
exhibit oscillations in the THz region. A significant amount of
inter-band current can be present in a staggered band-gap
structure. The coupling between the conduction and the valence
bands is considered to be the dominant mechanism in these
structures. Therefore, in a correct description of electron
transport, multi-band effects must be included \cite{key-6}. We do
not consider an external magnetic field but the extension is quite
straightforward.

For direct band-gap semiconductors the conduction band near $k=0$
has symmetry properties (spherical symmetry) same as the $|S>$ atomic
orbital ($l=0,m_{l}=0$). On the other hand, the valence band near
$k=0$ has symmetry of p-orbitals, $|X>,|Y>$and $|Z>$ (p-orbitals
are antisymmetric and $l=1,m_{l}=-1,0,1$). An eight-band model \cite{key-7}
can be gained with the inclusion of spin. These states that become
doubly degenerate, $|S\uparrow>,|X\uparrow>,|Y\uparrow>,|Z\uparrow>,|S\downarrow>,|X\downarrow>,|Y\downarrow>,|Z\downarrow>$.
The spin-orbit interaction lifts the six-fold degeneracy of the valence
band and splits it into a four-fold degenerate and a two-fold degenerate
level. If the spin-orbit coupling is considered in the energy band
calculation of a semiconductor, the Bloch states become a mixture
of spin up and spin down states. This becomes important in asymmetric
quantum well devices. Inversion asymmetry of the bulk or the confining
potential causes spin splitting even in zero magnetic field due to
spin-orbit interaction. Therefore inclusion of spin and spin-orbit
interaction in the quantum transport equations is important.

Derivations of quantum transport equations have usually been based
on the first-order gradient expansion \cite{key-30}. This
approximation is based on the assumption that the {}``fast''
quantum variations can be separated from the {}``slow''
macroscopic variations and causes the loss of information related
to quantum processes such as interference and tunneling which are
crucial in nano-scale devices. Buot and Jensen \cite{key-25},
\cite{key-8} presented an alternative derivation for single-band
devices and provided an exact integral form of the quantum
transport equation which is capable of accurately describing full
quantum effects. The first-order gradient expansion is still
needed to simplify the collision terms after the derivation of
transport equations. Their approach has been generalized for the
multi-band transport in this work.

The Wigner function modeling of the quantum transport in the
single-band resonant tunneling structures has been quite popular
in the literature due to its success in dealing with the
dissipation and the open boundary conditions \cite{key-25}.
Similarly, it is expected that one should be able to model the
multi-band transport in the resonant tunneling structures using
the Wigner function.

There are a number of works on multi-band Wigner function
representation of quantum transport in the literature. Miller and
Neikirk \cite{key-9} used Wigner function for multi-subband
transport in double barrier resonant tunneling structures. Demei
et al. presented multi-band Wigner function formulation without
spin \cite{key-10}, \cite{key-11}. Zhao et al. \cite{key-12}
showed that multi-band quantum transport equations can be
decoupled to reduce the number equations to be solved. Borgioli
\cite{key-13} employed a two-band Kane model to derive Wigner
function equations for resonant inter-band tunneling diodes.

The point of this work is to develop a complete theory of the
multi-band Wigner function for transport in nano-scale devices.
This has been accomplished by using non-equilibrium Green's
function methodology which is known to be the most complete
description of quantum transport. The results give us the Wigner
function formulation of multi-band systems based on ${\bf k.p}$
theory. The results can be easily simplified by the symmetry
arguments of the band structure of the system under study. The
derived multi-band Wigner function equations which are also
capable of description of zero magnetic field spin transport
devices are the first in the literature.

In the introduction, the subsection 1.1 we present a preliminary
on $\bf{k.p}$ method. Then in the subsection 1.2, the
non-equilibrium Green's function method in phase space is given.
We derive the Wigner function equations for multi-band systems in
section 2. Finally, in part section 3, we apply the formalism to a
simple one dimensional two-band resonant inter-band tunneling
diode.

\subsection{$\bf{k.p}$ Hamiltonian}

The Schr\"{o}dinger equation for the lattice periodic part of the
Bloch functions can be written as \cite{key-24}

\begin{eqnarray}
[\frac{\hat{p}^{2}}{2m_{0}}+V({\bf
r})+\frac{\hbar^{2}k^{2}}{2m_{0}}+\frac{\hbar}{m_{0}}{\bf
k}.\hat{{\bf p}} +\frac{\hbar^{2}}{4m_{0}^{2}c^{2}}({\bf {\bf
\hat{{\bf \sigma}}}}\times\bigtriangledown V).\hat{{\bf
p}}\nonumber
\\+\frac{\hbar}{4m_{0}^{2}c^{2}}({\bf \hat{{\bf
\sigma}}}\times\bigtriangledown V).{\bf k}]|n{\bf
k}>=\varepsilon_{n{\bf k}}|n{\bf k}>.  \end{eqnarray} We express
the bulk band matrix element $H_{ab}$ of the Hamiltonian in the
second order $\bf{k.p}$ theory as

\begin{equation}
H_{ab}=D_{ab}^{(2)\alpha\gamma}k_{\alpha}k_{\gamma}+D_{ab}^{(1)\alpha}k_{\alpha}+(D_{aa}^{(0)}+V_{a}({\bf r}))\delta_{ab}\end{equation}
 where the indices $\alpha$ and $\gamma$ are summed over $x$, $y$,
and $z$. The $a$, $b$ include both the band and the spin indices.
$V({\bf r})$ is a spin-independent self-consistent potential. Note
that for heterostructures $k_{\alpha}$ is replaced by $-i{\bf \bigtriangledown}_{\alpha}$.

We define a vector $\pi$ as,

\begin{equation}
\mathbf{\pi}=\mathbf{p}+\frac{\hbar}{4m_{0}c^{2}}(\bf{\hat{\sigma}}\times\mathbf{\bigtriangledown}V).
\end{equation}
So,

\begin{equation}
D_{ab}^{(2)\alpha\gamma}=\frac{\hbar^{2}}{2m_{0}}\delta_{ab}\delta_{\alpha\gamma}+(\frac{\hbar}{m_{0}})^{2}\sum_{r}\frac{\pi_{ar}^{\alpha}\pi_{rb}^{\gamma}}{(\frac{E_{a}+E_{b}}{2}-E_{r})},\end{equation}
 noting that the second term arises from L\"{o}wdin renormalization
and needed to include the interactions with remote bands. So we denote
these states by the index $r$. These interactions are usually ignored
in Kane models and so $D_{ab}^{(2)\alpha\gamma}(a\neq b)$ terms vanish.

The part of $H_{ab}$ linear in $k$ include the inter-band coupling
($\bf{k.p}$ interaction) and the spin-orbit interaction terms,

\begin{equation}
D_{ab}^{(1)\alpha}=\frac{\hbar}{m_{0}}\pi_{ab}^{\alpha}=\frac{\hbar}{m_{0}}p_{ab}^{\alpha}+\frac{\hbar}{4m_{0}^{2}c^{2}}(\hat{\mathbf{\sigma}}\times\mathbf{\bigtriangledown}V)_{ab}^{\alpha}\end{equation}
 where

\begin{equation}
\pi_{ab}^{\alpha}=<U_{a}|\hat{p}^{\alpha}|U_{b}>+\frac{1}{4m_{0}c^{2}}<U_{a}|(\hat{\mathbf{\sigma}}\times{\bf \bigtriangledown}V)^{\alpha}|U_{b}>\end{equation}
for $a\neq b$. Note that $\pi_{ba}^{\alpha}=(\pi_{ab}^{\alpha})^{*}$
and $\pi_{aa}^{\alpha}=0$ (which implies that $D_{aa}^{(1)\alpha}$terms
vanish).

The terms $p_{ab}^{\alpha}=<U_{a}|\hat{p}^{\alpha}|U_{b}>$ are the
inter-band momentum matrix elements and measure the strength of the
coupling between the various bands. Note that $\pi_{aa}^{\alpha}=0$
even if the band minimum is at some point other than ${\bf k}=0$.
The term $\frac{\hbar}{m_{0}}p_{ab}^{\alpha}$ is usually written
in terms of a real parameter $P$ originally defined by Kane. The
value of this parameter is known for any given material.

\begin{equation}
P=-\frac{i\hbar}{m_{0}}<S|\hat{p}_{x}|X>=-\frac{i\hbar}{m_{0}}<S|\hat{p}_{y}|Y>=-\frac{i\hbar}{m_{0}}<S|\hat{p}_{z}|Z>.\end{equation}
The band edge is denoted by $D_{aa}^{(0)}$ such that, $D_{aa}^{(0)}=E_{a}(\mathbf{k}=0)$.
We write $D_{aa}^{(0)}+V_{a}({\bf r})$ as $E_{a}({\bf r})$ in the
calculations.

\subsection{The non-equilibrium Green's function formalism in phase space}

The multi-band Green's function \cite{key-2} is defined by

\begin{equation}
G_{ab}(1,2)=-\frac{i}{\hbar}<\psi_{a}(1)\psi_{b}^{\dagger}(2)>_{C}\end{equation}
 where C denotes that time arguments are on a contour rather than
a real-time axis. The expectation value is defined in a grand-canonical
ensemble. We define the space time arguments $1=({\bf \mathbf{r}_{1},t_{1})}$,
$2=({\bf {\bf \mathbf{r}}_{2},t_{2})}$. $\psi_{a}$ is the field
operator for electrons. The $a$,$b$ include both the band and the
spin indices. The equation of motion of the band electron Green's
function is written as (sum over repeated indices)

\begin{equation}
[i\hbar\delta_{a\beta}\frac{\partial}{\partial
t_{1}}-H_{a\beta}(1)]G_{\beta b}(1,2)=\delta_{ab}\delta(1-2)+\int
d3\Sigma_{a\beta}(1,3)G_{\beta b}(3,2),\end{equation} and the
adjoint equation is given by

\begin{equation}
[-i\hbar\delta_{a\beta}\frac{\partial}{\partial
t_{2}}-H_{a\beta}(2)]G_{\beta b}(1,2)=\delta_{ab}\delta(1-2)+\int
d3G_{a\beta}(1,3)\Sigma_{\beta b}(3,2),\end{equation} where the
self-energy is denoted by $\Sigma(1,2)$. It describes the
scattering of electrons by other electrons, phonons and
impurities. Throughout the paper, Greek indices are used to denote
the repeated indices to be summed over.

The generalized Kadanoff-Baym (GKB) equation describes the time evolution
of the electron correlation function $G_{aa}^{<}(1,2)$ in the band
$a$. It should be noted that, in the equal time limit ($t_{1}=t_{2}$),
the off-diagonal ($a\neq b$) lesser Green's functions correspond
to inter-band polarizations in energy band space and inter-spin-band
polarizations in spin space whereas the diagonal Green's functions
give the particle densities with the spin up or down in each band.
The time evolution of $G^{<}(1,2)$ given by the GKB equation can
be written, using the Langreth algebra, as\begin{eqnarray}
i\hbar(\frac{\partial}{\partial t_{1}}+\frac{\partial}{\partial t_{2}})G_{ab}^{<}(1,2) & = & [H,G^{<}](1,2)_{ab}+[\Sigma^{<},\textrm{Re}G^{R}](1,2)_{ab}\nonumber \\
 &  & +\frac{i}{2}\{\Sigma^{<},A\}(1,2)_{ab}-\frac{i}{2}\{\Gamma,G^{<}\}(1,2)_{ab}.\label{gkb1}\end{eqnarray}
where $[$ $]$ is the commutation and $\{$$\}$ is the anti-commutation.
Above, the spectral function is defined as,

\begin{equation}
A_{ab}(1,2)=i[G_{ab}^{>}(1,2)-G_{ab}^{<}(1,2)]=-2\textrm{Im}[G_{ab}^{R}(1,2)],\end{equation}
and the dissipation function is

\begin{equation}
\Gamma_{ab}(1,2)=i[\Sigma_{ab}^{>}-\Sigma_{ab}^{<}]=-2\textrm{Im}[\Sigma_{ab}^{R}(1,2)].\label{deneme}\end{equation}

Switching to center of mass and relative coordinates (Wigner coordinates)
done by defining

\begin{equation}
\mathbf{R}=\frac{\mathbf{{\bf
\mathbf{r}_{1}+\mathbf{r}_{2}}}}{2};T=\frac{
t_{1}+t_{2}}{2},\end{equation}
 \begin{equation}
\mathbf{v}={\mathbf r}_{2}-{\mathbf
r}_{1};t=t_{2}-t_{1}.\end{equation}

The four-dimensional, (3+1), crystal momentum and its conjugate variable
lattice coordinate are represented as $p=({\bf \mathbf{p}},E)$, $r=(\mathbf{R},T)$.
Note that we use $\hbar\mathbf{k}$ as the crystal momentum when the
matrix elements of the Hamiltonian is considered. Therefore $\hbar{\bf \mathbf{k}}$
and $\mathbf{p}$ are used as the crystal momentum interchangeably.

The Weyl-Wigner representation, $W[\hat{O}]=O(p,r)$, of any operator
$\hat{O}(1,2)$ \cite{key-14}, \cite{key-15} is defined by

\begin{equation}
O(p,r)=\int
dv\exp(\frac{i}{\hbar}p.v)<R-\frac{v}{2}|\hat{O}|R+\frac{v}{2}>.\end{equation}
It is very important to note that $O(p,r)$ is real if $\hat{O}$ is
hermitian. Let $\hat{C}=\hat{A}\hat{B}$ then the differential form
of the Weyl transform of the product of two operators can be
obtained by,

\begin{eqnarray}
W[\hat{C}]=C(p,r) & = & \exp(i\hat{\Lambda})\hat{A}(p,r)\hat{B}(p,r)\nonumber \\
 & = & \exp(-i\hat{\Lambda})\hat{B}(p,r)\hat{A}(p,r),\end{eqnarray}
where

\begin{equation}
\hat{\Lambda}=\frac{\hbar}{2}[\frac{\partial^{(A)}}{\partial r}\frac{\partial^{(B)}}{\partial p}-\frac{\partial^{(A)}}{\partial p}\frac{\partial^{(B)}}{\partial r}].\end{equation}
The partial differential $\partial^{(A)}$acts on only $A$ and $\partial^{(B)}$
acts on $B$ only.

To obtain the Wigner function equation, it is necessary to switch
to a phase space description of GKB. Taking the Weyl-Wigner transform
of both sides of the equation (\ref{gkb1}) gives the GKB in the phase-space-energy-time
domain

\begin{eqnarray}
i\hbar\frac{\partial}{\partial T}G_{ab}^{<}(p,r) & = & \exp(i\hat{\Lambda})[H,G^{<}](p,r)_{ab}+\exp(i\hat{\Lambda})[\Sigma^{<},ReG^{R}](p,r)_{ab}\nonumber \\
 &  & +\frac{i}{2}\exp(i\hat{\Lambda})\{\Sigma^{<},A\}(p,r)_{ab}-\frac{i}{2}\exp(i\hat{\Lambda})\{\Gamma,G^{<}\}(p,r)_{ab}\label{gkb2}\end{eqnarray}
For any operator $A$ and $B$, the integral representations of
 $\exp(i\hat{\Lambda})A(p,r)B(p,r)$ and
 $\exp(-i\hat{\Lambda})A(p,r)B(p,r)$ in (3+1) dimensions can be
written as \cite{key-2},

\begin{eqnarray}
\exp(\pm i\hat{\Lambda})A(p,r)B(p,r) & = & \frac{1}{(h^{4})^{2}}\int dr_{1}dp_{1}dr_{2}dp_{2}\exp[\frac{i}{\hbar}p_{1}.(r-r_{2})]\exp[\frac{i}{\hbar}r_{1}.(p-p_{2})]\nonumber \\
 &  & \times A(p\pm\frac{p_{1}}{2},r\mp\frac{r_{1}}{2})B(p_{2},r_{2}),\label{integ rep1}\end{eqnarray}
Defining

\begin{equation}
K_{A}^{\pm}(p,r-r_{2};r,p-p_{2})=\int dp_{1}dr_{1}\exp(\frac{i}{\hbar}p_{1}.(r-r_{2}))\exp(\frac{i}{\hbar}r_{1}.(p-p_{2}))A(p\pm\frac{p_{1}}{2},r\mp\frac{r_{1}}{2}),\end{equation}
 the equation (\ref{gkb2}) becomes

\begin{eqnarray}
i\hbar\frac{\partial}{\partial T}G_{ab}^{<}(p,r) & = & \frac{1}{(h^{4})^{2}}\int dp_{2}dr_{2}K_{H_{a\beta}}^{c}(p,r-r_{2};r,p-p_{2})G_{\beta b}^{<}(p_{2},r_{2})\nonumber \\
 &  & +\frac{1}{(h^{4})^{2}}\int dp_{2}dr_{2}K_{\Sigma_{a\beta}^{<}}^{c}(p,r-r_{2};r,p-p_{2})ReG_{\beta b}^{R}(p_{2},r_{2})\nonumber \\
 &  & +\frac{i}{2(h^{4})^{2}}\int dp_{2}dr_{2}K_{\Sigma_{a\beta}^{<}}^{s}(p,r-r_{2};q,p-p_{2})A_{\beta b}(p_{2},r_{2})\nonumber \\
 &  & -\frac{i}{2(h^{4})^{2}}\int dp_{2}dr_{2}K_{\Gamma_{a\beta}}^{s}(p,r-r_{2};r,p-p_{2})G_{\beta b}^{<}(p_{2},r_{2}),\label{gkb3}\end{eqnarray}
where $K_{A}^{s,c}(p,r-r_{2};r,p-p_{2})=K_{A}^{+}(p,r-r_{2};r,p-p_{2})\pm K_{A}^{-}(p,r-r_{2};r,p-p_{2}).$

The multi-band Wigner function is found by taking the energy integral
of the Weyl-Wigner transformed $G_{ab}^{<}$:

\begin{equation}
f_{ab}(\mathbf{p},\mathbf{R},T)=\int dE(-i)G_{ab}^{<}(\mathbf{p},E;\mathbf{R},T).\label{wigner1}\end{equation}
Note that the indices $a$, $b$ include both spin and band. The total
Wigner function of the multi-band system with spin can be written
as the summation over the band and the spin indices,

\begin{equation}
f(\mathbf{p},\mathbf{r},\kappa)=\sum_{c,d}\sum_{m,m'=\uparrow,\downarrow}\sigma_{m,m'}^{\kappa}f_{cd}^{mm'}(\mathbf{p},\mathbf{r}),\end{equation}
where $c$, $d$ are band, $m$ and $m^{'}$ are spin indices.
$\kappa$ takes values of $x$, $y$, and $z$. $\sigma^{0}$ is the
unit matrix and the others are the Pauli spin matrices
\cite{key-16}. Therefore each intra-band and inter-band component
of Wigner function becomes $2\times2$ matrix in spin space.

\section{The Wigner Function Equations for Multi-band Systems}

Under the assumption that the self-energies are slowly varying with
respect to the center of mass coordinates, equation (\ref{gkb3})
reduces to

\begin{eqnarray}
i\hbar\frac{\partial}{\partial T}G_{ab}^{<}(p,r) & = & \frac{1}{(h^{4})^{2}}\int dp_{2}dr_{2}K_{H_{a\beta}}^{c}(p,r-r_{2};r,p-p_{2})G_{\beta b}^{<}(p_{2},r_{2})\nonumber \\
 &  & +\Sigma_{a\beta}^{>}(p,r)G_{\beta b}^{<}(p,r)-\Sigma_{a\beta}^{<}(p,r)G_{\beta b}^{>}(p,r)\label{eq:0.2.1}\end{eqnarray}
using

\begin{equation}
i[\Sigma_{a\beta}^{<}A_{\beta b}-\Gamma_{a\beta}G_{\beta b}^{<}]=\Sigma_{a\beta}^{>}G_{\beta b}^{<}-\Sigma_{a\beta}^{<}G_{\beta b}^{>}.\end{equation}
The self-energy function can be written as \cite{key-17},

\begin{equation}
\Sigma_{ab}^{>,<}(1,2)=iG_{ab}^{>,<}(1,2)D^{>,<}(1,2).\label{selfenergy1}\end{equation}
The Weyl-Wigner transform gives of the above equation (\ref{selfenergy1})
gives,

\begin{equation}
\Sigma_{ab}^{>,<}(p,r)=\frac{i}{h^{4}}\int dqG_{ab}^{>,<}(p+q)D^{>,<}(q).\end{equation}
Assuming the phonon bath is in equilibrium, the Fourier transforms
of the phonon Green's functions can be written as,

\begin{equation}
D^{<}(\mathbf{q},E^{'})=-ihM_{\mathbf{q}}^{2}[(N_{\mathbf{q}}+1)\delta(E^{'}-\Omega_{\mathbf{q}})+N_{\mathbf{q}}\delta(E^{'}+\Omega_{\mathbf{q}})],\end{equation}

\begin{equation}
D^{>}(\mathbf{q},E^{'})=-ihM_{\mathbf{q}}^{2}[(N_{\mathbf{q}}+1)\delta(E^{'}+\Omega_{\mathbf{q}})+N_{\mathbf{q}}\delta(E^{'}-\Omega_{\mathbf{q}})]\end{equation}
where $M_{\mathbf{q}}$ is the electron-phonon scattering matrix element.
Therefore, inclusion of the phonon scattering gives the following
scattering functions

\begin{equation}
\Sigma_{ab}^{<}=\sum_{\eta=+1,-1}\frac{1}{h^{3}}\int d{\bf q}G_{ab}^{<}(\mathbf{p}+\mathbf{q},E+\eta\Omega_{\mathbf{q}},r)M_{\mathbf{q}}^{2}(N_{\mathbf{q}}+\frac{1}{2}+\frac{1}{2}\eta),\end{equation}

\begin{equation}
\Sigma_{ab}^{>}=\sum_{\eta=+1,-1}\frac{1}{h^{3}}\int d{\bf q}G_{ab}^{>}(\mathbf{p}+\mathbf{q},E+\eta\Omega_{\mathbf{q}},r)M_{\mathbf{q}}^{2}(N_{{\bf \mathbf{q}}}+\frac{1}{2}-\frac{1}{2}\eta).\end{equation}
The first term on the right hand side of the equation (\ref{eq:0.2.1})
can be written as

\begin{equation}
exp(i\Lambda)[H,G^{<}](p,r)_{ab}=exp(i\Lambda)H_{a\beta}(p,r)G_{\beta b}^{<}(p,r)-exp(-i\Lambda)H_{\beta b}(p,r)G_{a\beta}^{<}(p,r).\end{equation}
 The integral representation, using the equation (\ref{integ rep1})
becomes

\begin{eqnarray}
\exp(i\Lambda)[H,G^{<}](p,r)_{ab} & = & \frac{1}{(h^{4})^{2}}\int dr_{1}dp_{1}dr_{2}dp_{2}\exp(\frac{i}{\hbar}p_{1}.(r-r_{2}))\exp(\frac{i}{\hbar}r_{1}.(p-p_{2}))\nonumber \\
 &  & \times[H_{a\beta}({\bf p}+\frac{{\bf p}_{1}}{2},{\bf q}-\frac{{\bf q}_{1}}{2})G_{\beta b}^{<}(p_{2},q_{2})\nonumber \\
 &  & -H_{\beta b}({\bf p}-\frac{{\bf p}_{1}}{2},{\bf r}+\frac{{\bf r_{1}}}{2})G_{a\beta}^{<}(p_{2},r_{2})],\end{eqnarray}
where

\begin{eqnarray}
H_{ab}(p\pm\frac{p_{1}}{2},r\mp\frac{r_{1}}{2}) & = & D_{ab}^{(2)\alpha\gamma}(p_{\alpha}\pm\frac{p_{1\alpha}}{2})(p_{\gamma}\pm\frac{p_{1\gamma}}{2})\nonumber \\
 &  & +D_{ab}^{(1)\alpha}(p_{\alpha}\pm\frac{p_{1\alpha}}{2})+(D_{aa}^{(0)}+V_{a}(\mathbf{r}\mp\frac{\mathbf{r}_{1}}{2}))\delta_{ab}.\end{eqnarray}

At this point, since the purpose of the paper is to derive a Boltzmann
type transport equation, it is useful to make quasiparticle approximation
to get the form of the spectral function. The free generalized Kadanoff-Baym
(FGKB) ansatz for multi-band systems is stated as \cite{key-18},

\begin{equation}
-iG_{ab}^{<}(\mathbf{p},E,\mathbf{r},T)=hf_{ab}(\mathbf{p},\mathbf{r},T)\delta(E-\frac{E_{a}(\mathbf{p})+E_{b}({\bf \mathbf{p}})}{2})\end{equation}

\begin{equation}
iG_{ab}^{>}(\mathbf{p},E,\mathbf{r},T)=h(\delta_{ab}-f_{ab}(\mathbf{p},\mathbf{r},T))\delta(E-\frac{E_{a}(\mathbf{p})+E_{b}(\mathbf{p})}{2})\end{equation}
Using FGKB ansatz, one can simplify the scattering functions so
that the equation of motion for $G^{<}$ in the
phase-space-energy-time domain becomes

\begin{eqnarray}
i\hbar\frac{\partial}{\partial T}G_{ab}^{<}(\mathbf{p},E,\mathbf{r},T) & = & D_{a\beta}^{(2)\alpha\gamma}p_{\alpha}p_{\gamma}G_{\beta b}^{<}-D_{\beta b}^{(2)\alpha\gamma}p_{\alpha}p_{\gamma}G_{a\beta}^{<}\nonumber \\
 &  & +\frac{\hbar}{2i}D_{a\beta}^{(2)\alpha\gamma}[p_{\alpha}\frac{\partial}{\partial r_{\gamma}}+p_{\gamma}\frac{\partial}{\partial r_{\alpha}}]G_{\beta b}^{<}+\frac{\hbar}{2i}D_{\beta b}^{(2)\alpha\gamma}[p_{\alpha}\frac{\partial}{\partial r_{\gamma}}+p_{\gamma}\frac{\partial}{\partial r_{\alpha}}]G_{a\beta}^{<}\nonumber \\
 &  & -\frac{\hbar^{2}}{4}D_{a\beta}^{(2)\alpha\gamma}\frac{\partial}{\partial r_{\alpha}}\frac{\partial}{\partial r_{\gamma}}G_{\beta b}^{<}+\frac{\hbar^{2}}{4}D_{\beta b}^{(2)\alpha\gamma}\frac{\partial}{\partial r_{\alpha}}\frac{\partial}{\partial r_{\gamma}}G_{a\beta}^{<}\nonumber \\
 &  & +D_{a\beta}^{(1)\alpha}[p_{\alpha}+\frac{\hbar}{2i}\frac{\partial}{\partial r_{\alpha}}]G_{\beta b}^{<}-D_{\beta b}^{(1)\alpha}[p_{\alpha}-\frac{\hbar}{2i}\frac{\partial}{\partial r_{\alpha}}]G_{a\beta}^{<}\nonumber \\
 &  & +\delta_{a\beta}\frac{1}{h^{3}}\int dp_{2}dr_{1}\exp(\frac{i}{\hbar}r_{1}.(p-p_{2}))[D_{aa}^{(0)}+V_{aa}({\bf r}-\frac{{\bf r}_{1}}{2})]G_{\beta b}^{<}(p_{2},r)\nonumber \\
 &  & -\delta_{\beta b}\frac{1}{h^{3}}\int dp_{2}dr_{1}\exp(\frac{i}{\hbar}r_{1}.(p-p_{2}))[D_{\beta\beta}^{(0)}+V_{\beta\beta}({\bf r}+\frac{{\bf r}_{1}}{2})]G_{a\beta}^{<}(p_{2},r)\nonumber \\
 &  & +[\Sigma_{a\beta}^{>}G_{\beta b}^{<}-\Sigma_{a\beta}^{<}G_{\beta b}^{>}]\end{eqnarray}
where $p_{\alpha}$ denotes $p_{x},p_{y},p_{z}$ , $\frac{\partial}{\partial r_{\alpha}}$
denotes $\frac{\partial}{\partial x},\frac{\partial}{\partial y},\frac{\partial}{\partial z}$
and there is summation over Greek indices as usual. Using the equation
(\ref{wigner1}), the multi-band Wigner functions for an open system
in weakly contact with a phonon heat bath can be written as

\begin{eqnarray}
i\hbar\frac{\partial}{\partial T}f_{ab}(\mathbf{p},\mathbf{r},T) & = & D_{a\beta}^{(2)\alpha\gamma}p_{\alpha}p_{\gamma}f_{\beta b}(\mathbf{p},\mathbf{r},T)-D_{\beta b}^{(2)\alpha\gamma}p_{\alpha}p_{\gamma}f_{a\beta}(\mathbf{p},\mathbf{r},T)\nonumber \\
 &  & +\frac{\hbar}{2i}D_{a\beta}^{(2)\alpha\gamma}[p_{\alpha}\frac{\partial}{\partial r_{\gamma}}+p_{\gamma}\frac{\partial}{\partial r_{\alpha}}]f_{\beta b}(\mathbf{p},\mathbf{r},T)\nonumber \\
 &  & +\frac{\hbar}{2i}D_{\beta b}^{(2)\alpha\gamma}[p_{\alpha}\frac{\partial}{\partial r_{\gamma}}+p_{\gamma}\frac{\partial}{\partial r_{\alpha}}]f_{a\beta}(\mathbf{p},\mathbf{r},T)\nonumber \\
 &  & -\frac{\hbar^{2}}{4}D_{a\beta}^{(2)\alpha\gamma}\frac{\partial}{\partial r_{\alpha}}\frac{\partial}{\partial r_{\gamma}}f_{\beta b}(\mathbf{p},\mathbf{r},T)+\frac{\hbar^{2}}{4}D_{\beta b}^{(2)\alpha\gamma}\frac{\partial}{\partial r_{\alpha}}\frac{\partial}{\partial r_{\gamma}}f_{a\beta}(\mathbf{p},\mathbf{r},T)\nonumber \\
 &  & +D_{a\beta}^{(1)\alpha}[p_{\alpha}+\frac{\hbar}{2i}\frac{\partial}{\partial r_{\alpha}}]f_{\beta b}(\mathbf{p},\mathbf{r},T)-D_{\beta b}^{(1)\alpha}[p_{\alpha}-\frac{\hbar}{2i}\frac{\partial}{\partial r_{\alpha}}]f_{a\beta}(\mathbf{p},\mathbf{r},T)\nonumber \\
 &  & +\delta_{a\beta}\frac{1}{h^{3}}\int dp_{2}dr_{1}\exp(\frac{i}{\hbar}r_{1}.(p-p_{2}))[D_{aa}^{(0)}+V_{aa}({\bf r}-\frac{{\bf r_{1}}}{2})]f_{\beta b}(\mathbf{p},\mathbf{r},T)\nonumber \\
 &  & -\delta_{\beta b}\frac{1}{h^{3}}\int dp_{2}dr_{1}\exp(\frac{i}{\hbar}r_{1}.(p-p_{2}))[D_{\beta\beta}^{(0)}+V_{\beta\beta}({\bf r}+\frac{{\bf r_{1}}}{2})]f_{a\beta}(\mathbf{p},\mathbf{r},T)\nonumber \\
 &  & +\sum_{\eta=+1,-1}\frac{1}{h^{3}}\int d\mathbf{q}(\delta_{a\beta}-f_{a\beta}(\mathbf{p}+\mathbf{q},\mathbf{r},T))f_{\beta b}({\bf \mathbf{p}},\mathbf{r},T)\nonumber \\
 &  & \times\delta(\frac{E_{a}({\bf \mathbf{p}}+{\bf \mathbf{q}})+E_{\beta}({\bf \mathbf{p}}+{\bf \mathbf{q}})}{2}-\frac{E_{\beta}({\bf \mathbf{p}})+E_{b}(\mathbf{p})}{2}+\eta\Omega_{\mathbf{q}})M_{\mathbf{q}}^{2}(N_{\mathbf{q}}+\frac{1}{2}-\frac{1}{2}\eta)\nonumber \\
 &  & -\sum_{\eta=+1,-1}\frac{1}{h^{3}}\int d\mathbf{q}f_{a\beta}(\mathbf{{\bf \mathbf{p}}+{\bf \mathbf{q}}},\mathbf{r},T)(\delta_{\beta b}-f_{\beta b}({\bf \mathbf{p}},\mathbf{r},T))\label{multibandwigner}\\
 &  & \times\delta(\frac{E_{a}({\bf \mathbf{p}}+{\bf \mathbf{q}})+E_{\beta}({\bf \mathbf{p}}+{\bf \mathbf{q}})}{2}-\frac{E_{\beta}({\bf \mathbf{p}})+E_{b}(\mathbf{p})}{2}+\eta\Omega_{\mathbf{q}})M_{\mathbf{q}}^{2}(N_{\mathbf{q}}+\frac{1}{2}+\frac{1}{2}\eta)\nonumber \end{eqnarray}
The third and the fourth terms at the right of the equation
(\ref{multibandwigner}) are the usual drift terms. The first, the
second, the fifth and the sixth terms do not cancel each other for
$a\neq b$ if and only if L\"{o}wdin renormalization is considered.
If the effects of the remote bands are ignored, these terms cancel
each other. The seventh and eight terms explicitly give the
$\mathbf{k.p}$ and spin-orbit interactions. The ninth and tenth
terms give the potential term. The last two terms correspond to
electron-phonon scattering. The relaxation time approximation can
be made for these. An important simplification occurs when the
in-plane wave vector is taken to be zero. This approximation gives
a set of spin-independent quantum transport equations.

If the structure under consideration has inversion symmetry as in
diamond structures, the equations can be simplified further. Note
that there is no inversion symmetry in zinc-blende structures (bulk
inversion asymmetry) and spin degeneracy in zinc-blende type heterostructures
is lifted even at zero magnetic field. Usually, this splitting is
very small and can be ignored. However, recently resonant intra-band
and inter-band spin filter was proposed based on the zero magnetic
field spin splitting of the conduction band in the case of structural
inversion asymmetry (Rashba effect) \cite{key-19}, \cite{key-20}.
We are going to discuss the Wigner function modeling of these kind
of devices in future papers.

The total current density is the sum of intra-band and inter-band
components \cite{key-14},

\begin{equation}
J(r)=\frac{q}{h^{3}}\int d\mathbf{p}\frac{\partial H_{\alpha\beta}}{\partial\mathbf{p}}f_{\beta\alpha}(\mathbf{p},r).\label{current}\end{equation}

The particle density in each band is written as

\begin{equation}
n_{a}=\frac{1}{h^{3}}\int d\mathbf{p}f_{aa}(\mathbf{p},r).\label{particle}\end{equation}

Let's consider a two-band model ($a,b=1,2$) without scattering and
neglect the effects of the remote bands. Then the first component
of the Wigner equation (\ref{multibandwigner}) becomes

\begin{eqnarray}
i\hbar\frac{\partial f_{11}(\mathbf{p},r)}{\partial T} & = & \frac{\hbar}{i}D_{11}^{(2)\alpha\gamma}[p_{\alpha}\frac{\partial}{\partial r_{\gamma}}+p_{\gamma}\frac{\partial}{\partial r_{\alpha}}]f_{11}(\mathbf{p},r)\nonumber \\
 &  & +D_{12}^{(1)\alpha\gamma}[p_{\alpha}+\frac{\hbar}{2i}\frac{\partial}{\partial r_{\alpha}}]f_{21}(\mathbf{p},r)-D_{21}^{(1)\alpha\gamma}[p_{\alpha}-\frac{\hbar}{2i}\frac{\partial}{\partial r_{\alpha}}]f_{12}(\mathbf{p},r)\nonumber \\
 &  & +\frac{1}{h^{3}}\int dp_{2}dr_{1}\exp(\frac{i}{\hbar}r_{1}.(p-p_{2}))\nonumber \\
 &  & \times[V_{1}(\textrm{{\bf r}}-\frac{{\bf r}_{1}}{2})-V_{1}({\bf r}+\frac{{\bf r}_{1}}{2})]f_{11}(\mathbf{p}_{2},r).\end{eqnarray}
The rest of the equations are as follows

\begin{eqnarray}
i\hbar\frac{\partial f_{12}(\mathbf{p},r)}{\partial T} & = & \frac{\hbar}{2i}[D_{11}^{(2)\alpha\gamma}+D_{22}^{(2)\alpha\gamma}][p_{\alpha}\frac{\partial}{\partial r_{\gamma}}+p_{\gamma}\frac{\partial}{\partial r_{\alpha}}]f_{12}(\mathbf{p},r)\nonumber \\
 &  & +D_{12}^{(1)\alpha}[p_{\alpha}+\frac{\hbar}{2i}\frac{\partial}{\partial r_{\alpha}}]f_{22}(\mathbf{p},r)-D_{12}^{(1)\alpha}[p_{\alpha}-\frac{\hbar}{2i}\frac{\partial}{\partial r_{\alpha}}]f_{11}(\mathbf{p},r)\nonumber \\
 &  & +\frac{1}{h^{3}}\int dp_{2}dr_{1}\exp(\frac{i}{\hbar}r_{1}.(p-p_{2}))\nonumber \\
 &  & \times[V_{1}(\textrm{{\bf r}}-\frac{{\bf r}_{1}}{2})-V_{2}({\bf r}+\frac{{\bf r}_{1}}{2})]f_{12}(\mathbf{p}_{2},r),\end{eqnarray}

\begin{eqnarray}
i\hbar\frac{\partial f_{22}(\mathbf{p},\mathbf{r},T)}{\partial T} & = & \frac{\hbar}{i}D_{22}^{(2)\alpha\gamma}[p_{\alpha}\frac{\partial}{\partial r_{\gamma}}+p_{\gamma}\frac{\partial}{\partial r_{\alpha}}]f_{22}(\mathbf{p},r)\nonumber \\
 &  & +D_{21}^{(1)\alpha\gamma}[p_{\alpha}+\frac{\hbar}{2i}\frac{\partial}{\partial r_{\alpha}}]f_{12}(\mathbf{p},r)-D_{12}^{(1)\alpha\gamma}[p_{\alpha}-\frac{\hbar}{2i}\frac{\partial}{\partial r_{\alpha}}]f_{21}(\mathbf{p},r)\nonumber \\
 &  & +\frac{1}{h^{3}}\int dp_{2}dr_{1}\exp(\frac{i}{\hbar}r_{1}.(p-p_{2}))\nonumber \\
 &  & \times[V_{2}(\textrm{{\bf r}}-\frac{{\bf r}_{1}}{2})-V_{2}({\bf r}+\frac{{\bf r}_{1}}{2})]f_{22}(\mathbf{p}_{2},r),\end{eqnarray}
and $f_{21}=f_{12}^{*}$.

\section{1-Dimensional Two-band Kane Model of Resonant Inter-band Tunneling
Diode}

Resonant inter-band tunneling structures (RITS) are based on the
interaction between the conduction and valence bands and the
transport is in the growth direction. For narrow band-gap RIT
structures, the coupling of the in-plane momentum to the
transverse momentum component becomes important. Therefore a
realistic modeling of these structures requires a serious amount
of computational work.

The simplest choice is a two-band model that is suitable for large
and mid-band-gap Type I RITS \cite{key-21}. The $\mathbf{k}$
vector is taken in the $z$ direction and the inversion asymmetry
is neglected. Therefore, the in-plane momentum ($k_{x}=k_{y}=0$)
vanishes so that the heavy-hole state is decoupled and the
Hamiltonian matrix becomes block-diagonal \cite{key-22}. The
remote bands are neglected too. Therefore the Hamiltonian is
reduced to a spin-independent three-band model (conduction, light
and split-off bands) \cite{key-23}:

\begin{equation}
\left[\begin{array}{ccc}
E_{c}(z)+\frac{p_{z}^{2}}{2m_{0}} & \sqrt{\frac{2}{3}}\frac{P_{cv}}{m_{0}}p_{z} & -\sqrt{\frac{1}{3}}\frac{P_{cv}}{m_{0}}p_{z}\\
-\sqrt{\frac{2}{3}}\frac{P_{cv}}{m_{0}}p_{z} & E_{lh}(z)+\frac{p_{z}^{2}}{2m_{0}} & 0\\
\sqrt{\frac{1}{3}}\frac{P_{cv}}{m_{0}}p_{z} & 0 &
E_{so}(z)+\frac{p_{z}^{2}}{2m_{0}}\end{array}\right].\end{equation}
where $P_{cv}=i\sqrt{\frac{m_{0}E_{p}}{2}}$. Instead of ignoring
the split-off band, Sirtori et. al. \cite{key-23} presented an
improved two-band model (conduction and {}``effective valence
band'') that can be gained through a unitary transformation. The
$2\times2$ Hamiltonian is

\begin{equation}
\left[\begin{array}{cc}
E_{c}(z)+\frac{p_{z}^{2}}{2m_{0}} & \frac{P_{cv}}{m_{0}}p_{z}\\
\frac{P_{cv}^{*}}{m_{0}}p_{z} & E_{v}(z)+\frac{p_{z}^{2}}{2m_{0}}\end{array}\right]\end{equation}
where $E_{v}=\frac{2E_{lh}+E_{so}}{3}$ is effective valence band
edge. Therefore the components of the Wigner function become

\begin{eqnarray}
i\hbar\frac{\partial f_{cc}(p_{z},z,t)}{\partial t} & = & -\frac{i\hbar p_{z}}{m_{0}}\frac{\partial f_{cc}(p_{z},z,t)}{\partial z}\nonumber \\
 &  & +\frac{1}{h}\int dz_{1}dp_{z2}\exp(\frac{i}{\hbar}z_{1}(p_{z}-p_{z2}))[E_{c}(z-\frac{z_{1}}{2})-E_{c}(z+\frac{z_{1}}{2})]f_{cc}(p_{z2},z,t)\nonumber \\
 &  & +\frac{p_{z}}{m_{0}}P_{cv}f_{vc}(p_{z},z,t)-\frac{i\hbar}{2m_{0}}P_{cv}\frac{\partial f_{vc}(p_{z},z,t)}{\partial z}\nonumber \\
 &  & +\frac{p_{z}}{m_{0}}P_{cv}f_{cv}(p_{z},z,t)+\frac{i\hbar}{2m_{0}}P_{cv}\frac{\partial f_{cv}(p_{z},z,t)}{\partial z},\end{eqnarray}

\begin{eqnarray}
i\hbar\frac{\partial f_{cv}(p_{z},z,t)}{\partial t} & = & -\frac{i\hbar p_{z}}{m_{0}}\frac{\partial f_{cv}(p_{z},z,t)}{\partial z}\nonumber \\
 &  & +\frac{1}{h}\int dz_{1}dp_{z2}\exp(\frac{i}{\hbar}z_{1}(p_{z}-p_{z2}))[E_{c}(z-\frac{z_{1}}{2})-E_{v}(z+\frac{z_{1}}{2})]f_{cv}(p_{z2},z,t)\nonumber \\
 &  & +\frac{p_{z}}{m_{0}}P_{cv}f_{vv}(p_{z},z,t)-\frac{i\hbar}{2m_{0}}P_{cv}\frac{\partial f_{vv}(p_{z},z,t)}{\partial z}\nonumber \\
 &  & +\frac{p_{z}}{m_{0}}P_{cv}f_{cc}(p_{z},z,t)-\frac{i\hbar}{2m_{0}}P_{cv}\frac{\partial f_{cc}(p_{z},z,t)}{\partial z},\end{eqnarray}

\begin{eqnarray}
i\hbar\frac{\partial f_{vv}(p_{z},z,t)}{\partial t} & = & -\frac{i\hbar p_{z}}{m_{0}}\frac{\partial f_{vv}(p_{z},z,t)}{\partial z}\nonumber \\
 &  & +\frac{1}{h}\int dz_{1}dp_{z2}\exp(\frac{i}{\hbar}z_{1}(p_{z}-p_{z2}))[E_{v}(z-\frac{z_{1}}{2})-E_{v}(z+\frac{z_{1}}{2})]f_{vv}(p_{z2},z,t)\nonumber \\
 &  & -\frac{p_{z}}{m_{0}}P_{cv}f_{cv}(p_{z},z,t)+\frac{i\hbar}{2m_{0}}P_{cv}\frac{\partial f_{cv}(p_{z},z,t)}{\partial z}\nonumber \\
 &  & -\frac{p_{z}}{m_{0}}P_{cv}f_{vc}(p_{z},z,t)-\frac{i\hbar}{2m_{0}}P_{cv}\frac{\partial f_{vc}(p_{z},z,t)}{\partial z},\end{eqnarray}
and $f_{vc}=f_{cv}^{*}$. Note that above equations are for the conduction
and valence band electrons and $E_{a}(z)=E_{a}(0)+V_{a}(z)$. The
current density can be calculated using the equation (\ref{current})
and given by

\begin{equation}
J=J_{intra}+J_{inter}=\frac{e}{h}\int
dp_{z}\frac{p_{z}}{m_{0}}(f_{cc}(p_{z})+f_{vv}(p_{z}))+\frac{2e}{h}\int
dp_{z}\sqrt{\frac{m_{0}E_{p}}{2}}\frac{1}{m_{0}}\textrm{Im}[f_{cv}(p_{z})].\end{equation}
where Im$[f_{cv}]$ means the imaginary part of $f_{cv}$. The
particle densities in each band using the equation
(\ref{particle}) are

\begin{equation}
n_{c}=\frac{1}{h}\int dp_{z}f_{cc}(p_{z}),\end{equation}

\begin{equation}
n_{v}=\frac{1}{h}\int dp_{z}f_{vv}(p_{z}).\end{equation}

\section{Conclusions}

In this paper we developed multi-band Wigner function formalism
including spin, which has a profound effect especially in narrow
band-gap semiconductors. We employed the $\bf{k.p}$ Hamiltonian to
derive the quantum transport equations for multi-band
semiconductors using non-equilibrium Green's function methodology
for systems weakly coupled to a phonon bath. It was shown that in
addition to drift, potential and scattering terms that exist in
single-band form of Wigner function, there are terms arising from
inter-band coupling and spin-orbit interaction. These terms are
source of the off-diagonal terms of Wigner function in energy band
space and spin space.

A two-band Kane model of resonant inter-band tunneling diode was
presented. The current and particle densities were derived for
this simple model. We are going to discuss the numerical solution
of the two-band and the three-band equations and present the
simulation results in future papers.

\section{Acknowledgments} This work was supported by a grant from
the Army Research Office's Defense University Research Initiative
on Nanotechnology (DURINT). We thank Greg Recine, Robert Murawski,
and Vadim Puller for helpful discussions.
%\end{acknowledgments}

\end{document}